\documentclass[fleqn,usenatbib]{mnras}

\usepackage{newtxtext,newtxmath}
\usepackage[T1]{fontenc}
\usepackage[utf8]{inputenc}

\DeclareRobustCommand{\VAN}[3]{#2}
\let\VANthebibliography\thebibliography
\def\thebibliography{\DeclareRobustCommand{\VAN}[3]{##3}\VANthebibliography}

\usepackage{graphicx}	% Including figure files
\usepackage{amsmath}	% Advanced maths commands
\usepackage{subcaption}
\title[Statistical Study of RBEs]{Statistical Study of Rapid Blue Excursions as Mass Conduits in Solar Atmosphere}

\author[G. Rahangdale et al.]{
Govindswaroop Rahangdale$^{1}$
\\
\\
$^{1}$ Independent Researcher}

\date{Accepted XXX. Received YYY; in original form ZZZ}

\pubyear{\the\year{}}

\begin{document}
\label{firstpage}
\pagerange{\pageref{firstpage}--\pageref{lastpage}}
\maketitle

% Abstract of the paper
\begin{abstract}
Rapid Blue Excursions (RBEs) are transient blue-shifted chromospheric absorption features widely interpreted as the on-disk counterparts of Type II solar spicules. We investigate their dynamic properties using high-cadence spectral observations combined with automated detection and spatio-temporal tracking algorithms. RBEs were identified through blue-wing Doppler asymmetry criteria and tracked using spatial connectivity and centroid continuity methods to determine lifetimes and kinematic evolution. The statistical analysis shows that RBEs are short-lived events with lifetimes predominantly between 20–60 s and a mean duration of approximately 75 s. The lifetime distribution follows an exponential decay profile, indicative of impulsive driving. Line-of-sight velocities range from 20 to 140 km s$^{-1}$, with a mean near 26 km s$^{-1}$. Projected lengths span 1.2–5.5 Mm with sub-arcsecond widths, and recurrence analysis reveals repeated activity at localized magnetic footpoints. Mass flux estimates suggest that RBEs transport significant plasma upward, contributing to chromospheric mass supply toward the corona. These findings reinforce the role of RBEs as dynamic conduits of mass transfer and key elements in chromosphere–corona coupling.
\end{abstract}

% Select between one and six entries from the list of approved keywords.
% Don't make up new ones.

%%%%%%%%%%%%%%%%%%%%%%%%%%%%%%%%%%%%%%%%%%%%%%%%%%

%%%%%%%%%%%%%%%%% BODY OF PAPER %%%%%%%%%%%%%%%%%%

\section{Introduction}

The solar chromosphere is a dynamic, complex layer filled with small-scale, short-lived bursts of plasma activity. Spicules—those thin, jet-like structures extending from the edge of the Sun—are some of its most prominent features. They shoot upward from the photosphere, often reaching heights of 10 Mm or more. Scientists have known about spicules since the late 19th century (\cite{1877arnp.book.....S}), and Beckers in the 1960s and 1970s laid the foundation for their spectroscopic investigation. Even today, spicules remain a mystery. Their true physical properties, how they are created, and their significance for the Sun’s atmosphere are still active areas of research.

With the advent of sharper, faster solar observations, we now recognize that spicules actually come in at least two distinct types. \cite{2007ApJ...655..624D} discovered a category of spicules—Type II—that are short-lived and move quickly, quite unlike the older, slower Type I spicules. Type I spicules last approximately 150 to 400 seconds and follow a well-defined arc, rising and falling with speeds between 15 and 40 km s$^{-1}$. These are well explained by magnetoacoustic shock waves (\cite{2006ApJ...647L..73H}). In contrast, Type II spicules are fleeting—lasting only 10 to 150 seconds—with higher velocities ranging from 30 to 110 km s$^{-1}$ (\cite{2012ApJ...759...18P}), and they typically do not show a clear return phase. Their rapid disappearance suggests they might be heating up to the much hotter transition region or even coronal temperatures (\cite{2007ApJ...655..624D}).

Type II spicules are most common in quiet Sun regions and coronal holes, while Type I spicules are mainly found in active regions (\cite{2007ApJ...655..624D}; \cite{2012ApJ...759...18P}). The ability to observe Type II spicules really made progress only with modern, high-speed, high-resolution telescopes. Older ground-based instruments simply could not capture these fast, ephemeral features (\cite{2012ApJ...759...18P}.

Observing spicules at the solar limb is challenging. There is so much overlap along the line of sight—a dense forest of jets—that following a single spicule is difficult. This crowding pushed researchers to look for their on-disk counterparts, which could be observed without as much confusion. \cite{2008ApJ...679L.167L} and \cite{2009ApJ...705..272R} made this connection: they identified sudden, transient blue shifts in chromospheric spectral lines—Rapid Blue-shifted Excursions, or RBEs—that matched the behavior of Type II spicules. Later studies by (\cite{2012ApJ...752..108S}; \cite{2013ApJ...769...44S}) developed a more complete picture of RBEs, reinforcing the association with Type II spicules. \cite{2012ApJ...759...18P} clarified their dynamic properties using extensive datasets.

\begin{figure}
\centering
\begin{subfigure}{0.8\columnwidth}
    \centering
    \includegraphics[width=\linewidth]{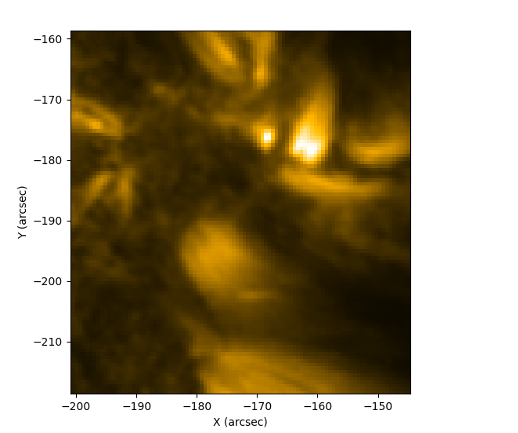}
    \caption{}
    \label{fig:aia171}
\end{subfigure}
\hfill
\begin{subfigure}{0.8\columnwidth}
    \centering
    \includegraphics[width=\linewidth]{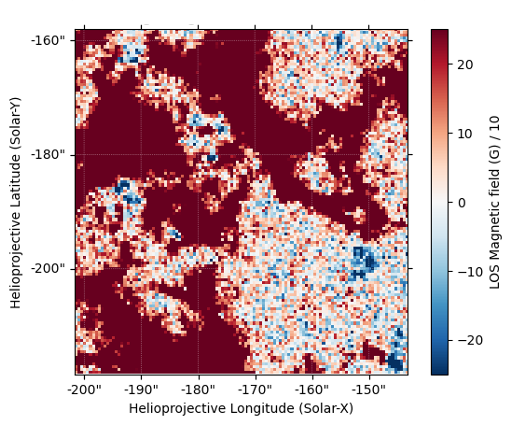}
    \caption{}
    \label{fig:hmi}
\end{subfigure}
\caption{
(a) SDO/AIA 171~$\mathring{A}$\ image of the observed region. (b) SDO/HMI line-of-sight magnetogram of the same field of view.
}
\label{fig:context_maps}
\end{figure}

Understanding Type II spicules and RBEs connects directly to the long-standing coronal heating problem—the question of why the Sun’s corona is at a million degrees while the surface below is much cooler. \cite{1943ApJ....98..116S} first demonstrated that the hot corona contains highly ionized atoms, proving the existence of extremely hot plasma. Since then, scientists have debated two major classes of heating mechanisms. Wave-based models, dating back to \cite{10.1093/mnras/107.2.211}, propose that magnetohydrodynamic waves transport energy upward and dissipate it in the corona (\cite{1961ApJ...134..347O}; \cite{1981JGR....8611463H}). The other approach favors reconnection-based models, where tiny explosions—nanoflares—release magnetic energy in short, impulsive events. Despite decades of research, it is still unclear which mechanism is most important.

Recently, chromospheric jets like spicules have been highlighted as possible links between the chromosphere and corona. Multi-instrument observations reveal that spicular plasma can change temperature as it rises, and some studies suggest it is heated to transition region or coronal temperatures (\cite{2009ApJ...701L...1D}; 2011). Early work by Beckers (1972) and \cite{1982ApJ...255..743A} estimated that spicules could send significant amounts of plasma upwards. More recently, even if only a fraction of a spicule is heated, it could still contribute to supplying mass to the corona (\cite{2009ApJ...701L...1D}). Additionally, researchers have detected transverse motions in spicules that resemble Alfvén waves (\cite{2007ApJ...655..624D}). \cite{2011ApJ...736L..24O} and \cite{2011Natur.475..477M} estimated the energy carried by these waves, indicating that spicules might channel magnetic energy directly into the upper solar atmosphere.

Radiative magnetohydrodynamic simulations that incorporate ambipolar diffusion and ion-neutral interactions have succeeded in reproducing jet-like eruptions resembling those observed in Type II spicules (\cite{2012ApJ...753..161M},\cite{2017Sci...356.1269M}). These models tie spicule formation to the release of magnetic tension in partially ionized plasma. They capture the key characteristics: rapid velocities, brief lifetimes, and swift temperature changes.

However, major questions remain. The details of how RBEs evolve thermodynamically, the extent of plasma heating during events, and the precise role these features play in shaping the upper atmosphere are all still not fully understood. To resolve these uncertainties, we require rapid, high-cadence spectroscopic observations as well as reliable detection and tracking methods.

In this study, we investigate the thermal evolution and behavior of on-disk Type II spicules using their RBE signatures. We utilize automated detection and tracking on high-resolution spectroscopic datasets, extracting statistics on their lifetimes, Doppler velocities, and spatial extents. Using Differential Emission Measure (DEM) analysis, we examine plasma temperature changes in regions where RBEs are found. By combining kinematic information with thermal diagnostics, our aim is to clarify how RBEs link the chromosphere and transition region, and to assess their true contribution to upper atmospheric heating—a topic first brought up by Edlén in 1943.

\section{Observations}

The observations were acquired using the Swedish 1-m Solar Telescope (SST; \cite{2003SPIE.4853..341S}) at La Palma. This was part of a joint campaign with the Interface Region Imaging Spectrograph (IRIS). For the imaging spectroscopic data, we used the CRisp Imaging SpectroPolarimeter (CRISP; \cite{2008ApJ...689L..69S}). CRISP relies on dual Fabry–Pérot interferometers (FPIs) and uses three synchronized high-speed, low-noise CCD cameras. These run at 35 frames per second, with each exposure lasting 17 milliseconds. The cameras sit behind the CRISP pre-filter and are synced up with an optical chopper, so all channels record data at the same time.
\begin{figure}
\centering
\includegraphics[width=0.9\columnwidth]{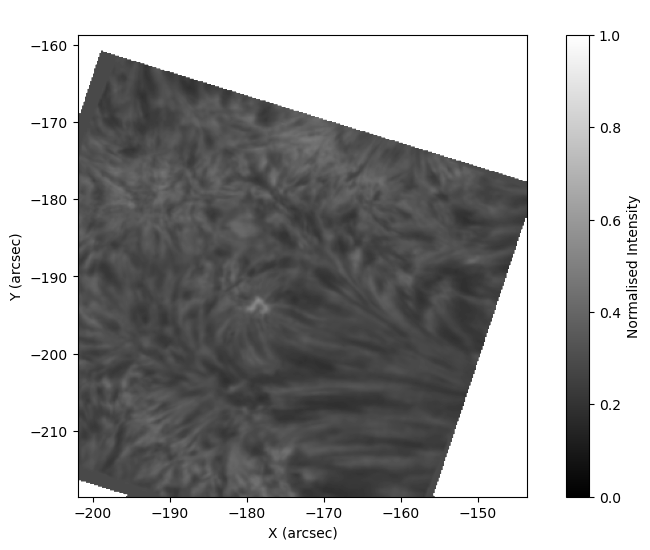}
\caption{
SST $H\alpha\ $core $6563$$\mathring{A}$\  intensity image of the observed Field of View, normalized to the background continuum intensity.}
\label{fig:halphacore}
\end{figure}
\begin{figure}
\centering
\includegraphics[width=0.9\columnwidth]{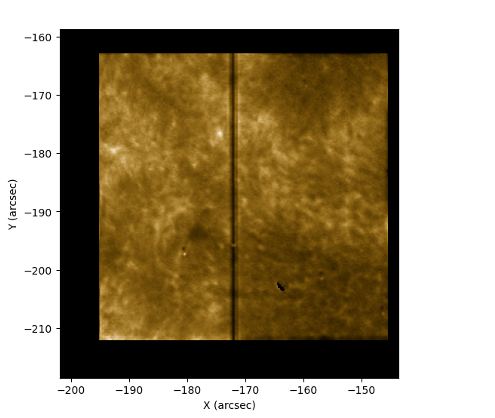}
\caption{
IRIS Mg\,\textsc{ii}\,k intensity image of the observed Field of View.}
\label{fig:irismg}
\end{figure}

The observations focus on the active region AR 11838 close to the disk center in the two-step sparse raster mode, at heliocentric coordinates ($x,y$) = ($-187''$, $-189''$). This spot corresponds to $\mu = \cos\theta = 0.70$. SST observed the region for about 73 minutes, from 08:25:52 UT to 09:34:37 UT, while IRIS covered a wider time window, from 08:09:43 UT to 11:09:43 UT. CRISP provided imaging spectroscopy in H$\alpha$ and collected Mg\,\textsc{ii}\,k spectra from IRIS. The CRISP wideband channel has a FWHM of 4.9\,$\mathring{A}$\ centered on H$\alpha$ - $6563 $$\mathring{A}$. For these observations, CRISP cycles through 24 different line positions, starting at 6561.8 $\mathring{A}$~ in the H$\alpha$ blue wing (that’s -1.2 $\mathring{A}$~ from line center) and moving step by step to 6564.2 $\mathring{A}$~ in the red wing (+1.2 $\mathring{A}$ ~ from line center). Each step is nearly 100 m$\mathring{A}$~ apart. For H$ \alpha$ 6563 $\mathring{A}$, CRISP delivers a transmission FWHM of 66 m$\mathring{A}$, while the pre-filter spans 4.9 $\mathring{A}$. To sharpen the time series images and remove leftover atmospheric distortions, we ran them through the Multi-Object Multi-Frame Blind Deconvolution (MOMFBD) image restoration method (\cite{2005SoPh..228..191V}). After processing, we analyzed the data with CRISPEX, a tool designed for handling multi-dimensional data cubes (\cite{2012ApJ...750...22V}). The H $\alpha$ images have a pixel size of 0.0592 arcsec, which offers about ten times the resolving power of SDO/AIA images (which are 0.6 arcsec per pixel). For the SDO/AIA images, we ran the standard SolarSoft reduction pipeline (\cite{2012SoPh..275...17L}). SDO/AIA observes the Sun in eight temperature channels, taking images every 12 seconds without interruption. Figure~\ref{fig:aia171} shows the SDO/AIA 171~$\mathring{A}$\ image of the observed region. The AIA reduction pipeline follows a set sequence: decompression and reversion, dead-time correction, local gain correction, flat-field correction, cosmic ray removal, and hot pixel correction. To reach sub-pixel accuracy when aligning AIA images with CRISP narrow-band data, we cross-correlated photospheric bright points in simultaneous CRISP wide-band images and AIA continuum (~5000 K) and EUV 171 $\mathring{A}$ images.

\section{Method}

\textit{RBE detection algorithm}
We have followed the below steps for the detection and tracking of RBEs (\cite{2022MNRAS.509.5523N}):
\begin{enumerate}
    \item The property of an event to be classified as an RBE depends on the line shift observed in the line profile of the event.
    \item The spectra is first normalized with respect to the continuum intensity. A reference profile was obtained by averaging the spectrum profile in a certain quiet sun region.
    \item Once the normalized spectrum is obtained, the algorithm scans through the obtained H${\alpha}$ spectrum profiles of each pixel throughout the entire field of view. This profile is then subtracted from the reference profile.
    \item This provides a residual profile for each pixel in the  FOV. Usually, these residual profiles have two peaks, one in the blue wing and one in the red wing, due to the effects of thermal line broadening phenomenon.
    \item The algorithm then finds the peak intensity value and the corresponding wavelength of these peaks by fitting a gaussian profile to the residual spectrum. By manual inspection, a threshold of 0.1 $I/<I>$ is applied to the spectrum profiles to properly identify RBEs and RREs.
    \item The pixel with the peak intensity value greater than the threshold in the blue wing is classified as an RBE, whereas the pixel with the peak intensity value greater than the threshold in the red wing is classified as an RRE. However, if a pixel has peak intensity value more than the threshold in both the blue and red wings, then it is classified as a broadening point. These points can not be classified as excursions on the account of line broadening. Values lower than 0.1 were also tried but this did not affect the number of RBEs and RREs.
\end{enumerate}

To quantify the spatial and temporal evolution of Rapid Blue Excursions (RBEs), we developed an automated tracking algorithm based on residual intensity enhancement, sub-pixel centroid estimation, and positional continuity constraints. For each time step, the algorithm scans along the horizontal slit to identify localized maxima in the residual peak intensity profile $I_{\mathrm{res}}(x,t)$. A local enhancement is defined as a region where the residual intensity exhibits a monotonic increase over at least five consecutive pixels followed by a monotonic decrease over at least five consecutive pixels. The pixel corresponding to the maximum residual intensity within this region is selected as the provisional RBE position. To improve positional accuracy beyond the discrete pixel resolution, we fit a Gaussian profile to the local intensity distribution surrounding the detected maximum:

    \begin{equation}
    I(x) = I_0 \exp\left[-\frac{(x - x_c)^2}{2\sigma^2}\right] + C,
    \end{equation}
where $I_0$ is the peak amplitude, $x_c$ is the centroid position, $\sigma$ is the standard deviation of the Gaussian profile, and $C$ represents a constant background level. The fitted centroid $x_c$ provides the RBE position with sub-pixel precision. To associate detections across consecutive frames, a positional continuity criterion is imposed. The projected transverse velocity limit is approximately 20 km s$^{-1}$ given the plate scale and cadence. If this condition is satisfied, detections are considered to belong to the same RBE event. Tests with a stricter threshold of two pixels ($\approx$ 10 km s$^{-1}$) resulted in the exclusion of higher-velocity RBEs, while larger thresholds increased the probability of incorrectly merging adjacent events. The adopted four-pixel criterion provides an optimal balance between sensitivity and event separation. To suppress noise-driven detections and short-lived artifacts, only events persisting for at least three consecutive time steps are retained.

\begin{figure}
\centering
\includegraphics[width=0.9\linewidth]{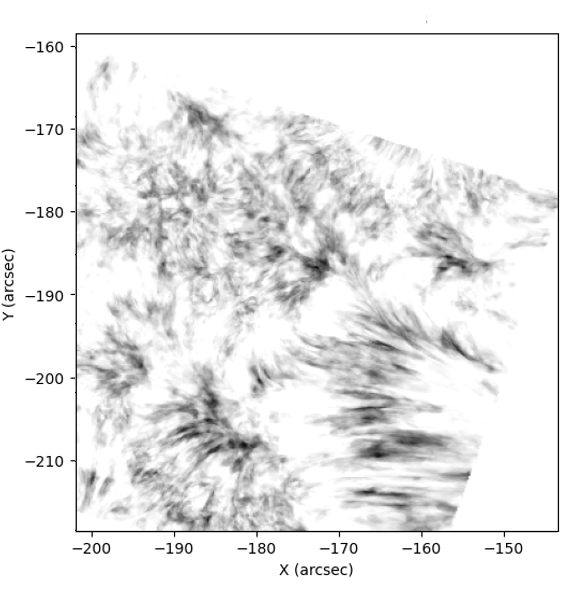}
\caption{Cumulative map showing all Rapid Blue Excursion (RBE) detections across the observing duration. Each dark feature corresponds to an identified RBE event. Darker regions indicate repeated activity.}
\label{fig:detection_map}
\end{figure}

\section{Results}

The automated detection pipeline applied to the high-cadence spectral dataset identified transient blue-wing absorption features consistent with Rapid Blue Excursions (RBEs). The detection algorithm was based on enhanced absorption in the blue wing of the H$\alpha$ profile relative to the line core and red wing intensity. A Doppler threshold criterion was applied to isolate significantly blue-shifted pixels, followed by spatial connectivity filtering to remove noise-dominated detections. Figure~\ref{fig:detection_map} shows a representative frame from the time sequence with detected RBEs overlaid on the chromospheric intensity map. The spatial distribution reveals clustering near magnetic network boundaries, consistent with magnetic driving mechanisms.
\begin{table}
\centering
\caption{Statistical properties of detected Rapid Blue Excursions (RBEs)}
\begin{tabular}{l c}
\hline
\hline
 & \textbf{Region} \\
\textbf{Properties} & \textbf{R1 (RBEs)} \\
\hline
Number per arcsec & 6.84 \\
Mean Lifetime (s) & $75s$ \\
Mean Doppler velocity (km\,s$^{-1}$) & $26.4$ \\
Mean projected length (Mm) & $3.1$ \\
\hline
\end{tabular}
\label{tab:rbe_statistics}
\end{table}

\begin{figure}
\centering
\includegraphics[width=0.8\linewidth]{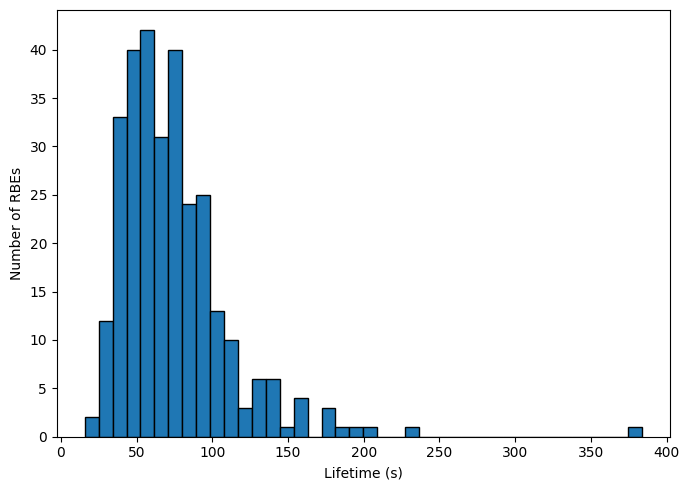}
\caption{Plot of the lifetimes of all the RBEs detected.}
\label{fig:lifetime_histogram}
\end{figure}

Randomly selected frames were checked for completeness of the detection through manual inspection. The algorithm successfully captures short-lived, elongated absorption structures while minimizing false positives from transient intensity fluctuations. Detected RBEs were tracked across consecutive frames using centroid continuity and spatial overlap criteria. Events were considered continuous if they exhibited non-zero pixel connectivity without temporal gaps exceeding one cadence interval. This ensured accurate lifetime estimation and prevented artificial splitting of extended structures. The resulting lifetime distribution is shown in Figure~\ref{fig:lifetime_histogram}. The majority of RBEs exhibit lifetimes between 20 and 200 seconds, with a mean value of approximately 75 seconds.

\begin{figure}
\centering
\includegraphics[width=0.8\linewidth]{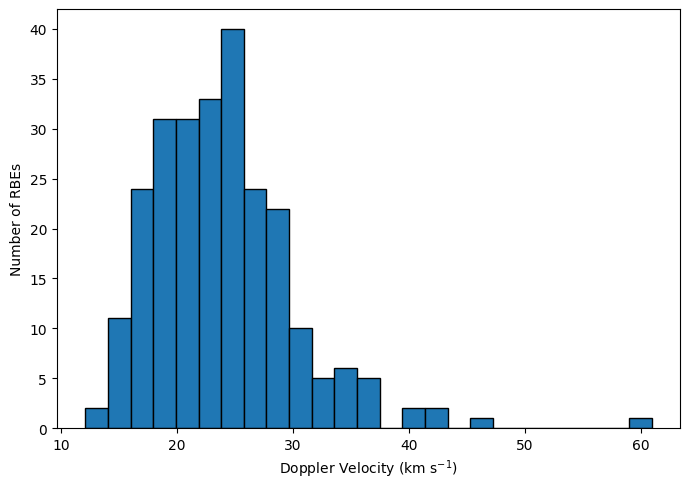}
\caption{Distribution of line-of-sight Doppler velocities for detected RBEs. The histogram shows high velocities characteristic of Type II spicules.}
\label{fig:velocity_histogram}
\end{figure}

The short lifetime regime supports the classification of these events as on-disk counterparts of Type II spicules. Line-of-sight velocities were derived from measured spectral shifts using,

\begin{equation}
v_{\mathrm{LOS}} = c \frac{\Delta \lambda}{\lambda_0},
\end{equation}

where $\Delta \lambda$ is the wavelength shift from line center and $\lambda_0$ is the rest wavelength. The velocity distribution is presented in Figure~\ref{fig:velocity_histogram}. Observed velocities span 20–140 km s$^{-1}$, with a mean velocity near 26 km s$^{-1}$. 

\begin{figure}
\centering
\includegraphics[width=0.8\linewidth]{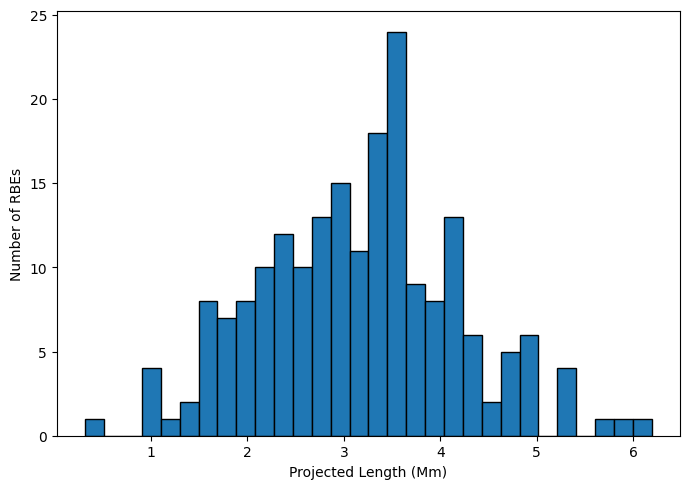}
\caption{Histogram of projected RBE lengths measured in megameters (Mm). The distribution peaks near 3 Mm.}
\label{fig:length_distribution}
\end{figure}

\begin{figure}
\centering
\includegraphics[width=0.8\linewidth]{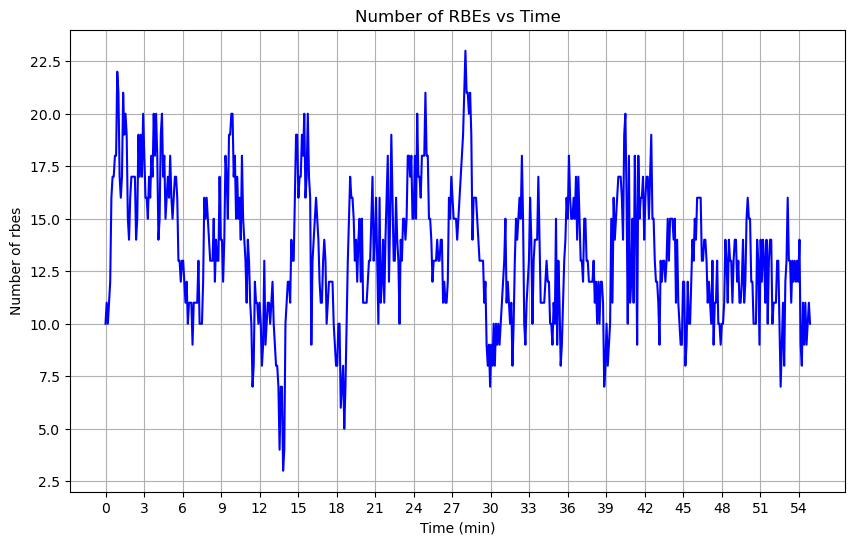}
\caption{Time evolution map of total number of RBE detections in the FOV across the observation period (in minutes).}
\label{fig:time_evolution}
\end{figure}

Temporal evolution analysis indicates that maximum velocities are typically reached during the early phase of the event lifetime, suggesting impulsive acceleration mechanisms. Projected lengths were calculated from connected pixel structures. RBEs exhibit lengths between 1.2 Mm and 5.5 Mm, with a mean value of 3.1 Mm. Widths remain confined to sub-arcsecond scales, typically below 300 km. Figure~\ref{fig:length_distribution} presents the distribution of projected lengths. Spatial recurrence maps might reveal repeated activity at specific magnetic footpoints. Figure~\ref{fig:time_evolution} shows the temporal evolution of the number of RBEs detected across the observation period, depicting variability in event occurance rate.

To study the thermal characteristics of plasma associated with RBEs and examine how these features might link the chromosphere and transition region. To do this, we performed a Differential Emission Measure (DEM) analysis, utilizing the SDO/AIA diagnostics. The DEM, $\mathrm{DEM}(T)$, indicates how plasma emitting at various temperatures is distributed and is defined as

\begin{equation}
\mathrm{DEM}(T) = n_e^2 \frac{dh}{dT},
\end{equation}

where, $n_e$ denotes the electron density, and $dh$ indicates the column depth of plasma within a particular temperature interval, $dT$. The observed intensity in a given spectral line or passband can be described by:

$G_\lambda(T)$, which is the temperature response or contribution function relevant to the diagnostic.

Assuming the plasma is optically thin and in thermal equilibrium, the DEM is directly related to the observed intensities through these temperature response functions, as described in \cite{2012A&A...539A.146H}. We determine these response functions using the CHIANTI atomic database (\cite{1997A&AS..125..149D}; \cite{1999ASSL..243...61L}). For the Mg\,\textsc{ii}\,k line, response functions are calculated under non-LTE conditions, utilizing precomputed contribution functions that are appropriate for chromospheric plasma.

To retrieve $\mathrm{DEM}(T)$ from the observed measurements, we applied the generalized regularized inversion technique outlined by \cite{2012A&A...539A.146H}, which is based on Tikhonov regularization (Tikhonov 1963). Because this inversion is an ill-posed problem, a smoothness constraint—following the approach of \cite{2004SoPh..225..293K} and \cite{2012A&A...539A.146H}—is imposed to ensure stability of the solution.

\begin{equation}
\chi^2 = \sum_i \frac{\left(I^{\mathrm{obs}}_i - I^{\mathrm{model}}_i\right)^2}{\sigma_i^2} + \lambda R[\mathrm{DEM}],
\end{equation}

where $\sigma_i$ represents the measurement uncertainties, $R$ imposes smoothness, and $\lambda$ serves as the regularization parameter—helping to stabilize the solution. The regularized inversion addresses the minimization problem directly and, through derivative estimation and smoothing, provides confidence intervals for each recovered temperature bin (see \cite{2012A&A...539A.146H}).

\begin{figure}
\centering
\includegraphics[width=0.9\linewidth]{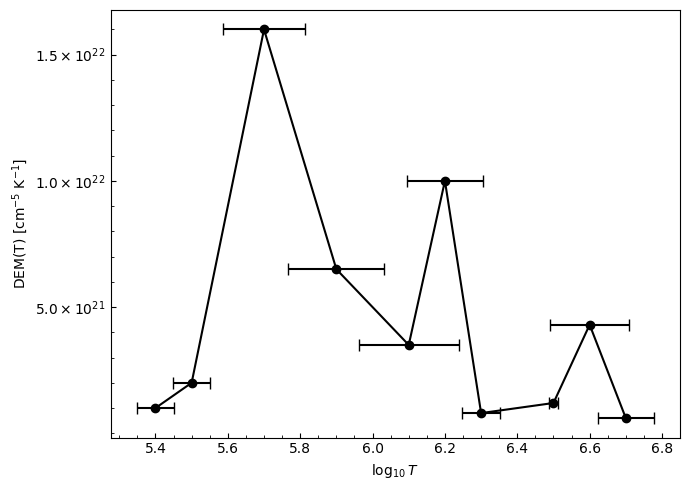}
\caption{Differential Emission Measure for a selected region.}
\label{fig:dem}
\end{figure}

We computed DEM solutions for spatial pixels corresponding to the tracked RBEs and compared them with adjacent quiet background regions. The resulting DEM distributions reveal a clear increase in emission between $\log T \approx 5.5$ and $6.2$ (K), which corresponds to the upper chromosphere and lower transition region. Pixels associated with RBEs consistently exhibit higher emission measure at temperatures above $\log T \sim 5.7$ compared to their surroundings. Figure~\ref{fig:dem} shows the DEM profile of a selected RBE.

The DEM profiles display a pronounced peak near upper chromospheric and lower coronal temperatures ($\log T \approx 5.7$), followed by a secondary rise extending toward $\log T \approx 6.2$. There is no significant emission detected above $\log T \gtrsim 6.8$, within the sensitivity limits of the observations.


\section{Conclusion and Discussion}

We set out to investigate the dynamic and thermal characteristics of on-disk Type II spicules—those Rapid Blue Excursions (RBEs) that stand out in high-cadence SST data, especially when combined with IRIS observations (\cite{2014ApJ...792L..15P}; \cite{2015ApJ...806..170S}). To do this, we developed an automated algorithm capable of detecting and tracking RBEs, allowing us to measure their motions, Doppler shifts, and overall statistical behavior (\cite{2009ApJ...705..272R}).

Most of the RBEs we analyzed persist for about 75 seconds, with some lasting up to 300 seconds, matching the previously reported lifetimes (\cite{2012ApJ...759...18P}; \cite{2013ApJ...769...44S}). Their Doppler velocities generally span 20 to 40 km s$^{-1}$, averaging around 24 km s$^{-1}$, and their typical length is about 3.1 Mm, showing agreement with \cite{2009ApJ...705..272R} and \cite{2013ApJ...764..164S}. To probe their thermal properties, we performed a Differential Emission Measure (DEM) analysis, focusing on pixels that corresponded to the tracked RBEs. The resulting DEM profiles show a strong signal at $\log T \approx 5.7$ corresponding to upper chromosphere and lower corona regions, along with a smaller enhancement extending to $\log T \approx 5.9 - 6.3$. We also observe that the DEM increases moderately during the most dynamic phase of the RBEs, suggesting localized heating or perhaps an increase in density as these jets move through the solar atmosphere (\cite{2014ApJ...792L..15P}; \cite{2015ApJ...806..170S}).

Altogether, this supports a scenario in which RBEs are more than just short-lived chromospheric features—they play an important role in connecting the chromosphere and transition region by moving plasma and inducing localized thermal changes (\cite{2008ApJ...679L.167L}; \cite{2014ApJ...792L..15P}). However, significant questions remain. How much of the RBE plasma actually undergoes substantial heating? How effectively do these processes convert energy? And how does the unresolved fine structure impact the DEM we observe? We also recognize that our analysis is limited, mainly by the temperature sensitivity and spectral coverage of current diagnostics. There is a possibility we are missing subtle coronal signals or rapid thermal transitions.

Future work, therefore, is required for better analysis by incorporating diagnostics sensitive to higher temperatures, improved inversion techniques, and coordinated observations from multiple instruments to clarify the thermodynamic evolution of RBEs. Advanced radiative MHD simulations—including realistic radiative transfer and ion-neutral interactions— can be used for connecting with the DEMs and kinematics we have measured. Combining these approaches can help determine whether RBEs genuinely play a role in heating the upper atmosphere or mainly reflect dynamic activity within the chromosphere. Sustaining high-resolution, fast-cadence spectroscopic observations with extended temperature coverage is essential to fully understand the role of chromospheric jets in the energetics of the solar atmosphere.

\section*{Acknowledgements}

The authors acknowledge the use of data from the Swedish 1-m Solar Telescope (SST), operated by the Institute for Solar Physics of Stockholm University at the Observatorio del Roque de los Muchachos, La Palma, Spain. We thank the IRIS team for open access to the data. IRIS is a NASA small explorer mission developed and operated by LMSAL with mission operations executed at NASA Ames Research Center and major contributions to downlink communications funded by ESA and the Norwegian Space Centre. SDO data are provided courtesy of the Joint Science Operations Center (JSOC) at Stanford University and the Instrument Operations Center (IOC) at Lockheed-Martin Solar and Astrophysics Laboratory (LMSAL). The SDO/AIA and SDO/HMI science teams are thanked for making the data publicly available. 

We also thank our colleagues, peers and collaborators for valuable discussions that improved the interpretation of the results and constructive feedback during the development of the methodology. This research has made use of the SolarSoft software package and community-developed analysis tools.
%%%%%%%%%%%%%%%%%%%%%%%%%%%%%%%%%%%%%%%%%%%%%%%%%%

%%%%%%%%%%%%%%%%%%%% REFERENCES %%%%%%%%%%%%%%%%%%

% The best way to enter references is to use BibTeX:

\bibliographystyle{mnras}
\bibliography{bibliography} % if your bibtex file is called example.bib

@BOOK{1877arnp.book.....S,
       author = {{Secchi}, Angelo},
        title = "{L'astronomia in Roma nel pontificato DI Pio IX.}",
         year = 1877,
       adsurl = {https://ui.adsabs.harvard.edu/abs/1877arnp.book.....S}
}

@ARTICLE{2007ApJ...655..624D,
       author = {{De Pontieu}, B. and {Hansteen}, V.~H. and {Rouppe van der Voort}, L. and {van Noort}, M. and {Carlsson}, M.},
        title = "{High-Resolution Observations and Modeling of Dynamic Fibrils}",
      journal = {\apj},
         year = 2007,
        month = jan,
       volume = {655},
       number = {1},
        pages = {624-641},
          doi = {10.1086/509070},
archivePrefix = {arXiv},
       eprint = {astro-ph/0701786},
 primaryClass = {astro-ph},
       adsurl = {https://ui.adsabs.harvard.edu/abs/2007ApJ...655..624D}
}

@ARTICLE{2006ApJ...647L..73H,
       author = {{Hansteen}, V.~H. and {De Pontieu}, B. and {Rouppe van der Voort}, L. and {van Noort}, M. and {Carlsson}, M.},
        title = "{Dynamic Fibrils Are Driven by Magnetoacoustic Shocks}",
      journal = {\apjl},
         year = 2006,
        month = aug,
       volume = {647},
       number = {1},
        pages = {L73-L76},
          doi = {10.1086/507452},
archivePrefix = {arXiv},
       eprint = {astro-ph/0607332},
 primaryClass = {astro-ph},
       adsurl = {https://ui.adsabs.harvard.edu/abs/2006ApJ...647L..73H}
}

@ARTICLE{2012ApJ...759...18P,
       author = {{Pereira}, Tiago M.~D. and {De Pontieu}, Bart and {Carlsson}, Mats},
        title = "{Quantifying Spicules}",
      journal = {\apj},
         year = 2012,
        month = nov,
       volume = {759},
       number = {1},
          eid = {18},
        pages = {18},
          doi = {10.1088/0004-637X/759/1/18},
archivePrefix = {arXiv},
       eprint = {1208.4404},
 primaryClass = {astro-ph.SR},
       adsurl = {https://ui.adsabs.harvard.edu/abs/2012ApJ...759...18P}
}

@ARTICLE{2008ApJ...679L.167L,
       author = {{Langangen}, {\O}. and {De Pontieu}, B. and {Carlsson}, M. and {Hansteen}, V.~H. and {Cauzzi}, G. and {Reardon}, K.},
        title = "{Search for High Velocities in the Disk Counterpart of Type II Spicules}",
      journal = {\apjl},
         year = 2008,
        month = jun,
       volume = {679},
       number = {2},
        pages = {L167},
          doi = {10.1086/589442},
archivePrefix = {arXiv},
       eprint = {0804.3256},
 primaryClass = {astro-ph},
       adsurl = {https://ui.adsabs.harvard.edu/abs/2008ApJ...679L.167L}
}

@ARTICLE{2009ApJ...705..272R,
       author = {{Rouppe van der Voort}, L. and {Leenaarts}, J. and {de Pontieu}, B. and {Carlsson}, M. and {Vissers}, G.},
        title = "{On-disk Counterparts of Type II Spicules in the Ca II 854.2 nm and H{\ensuremath{\alpha}} Lines}",
      journal = {\apj},
         year = 2009,
        month = nov,
       volume = {705},
       number = {1},
        pages = {272-284},
          doi = {10.1088/0004-637X/705/1/272},
archivePrefix = {arXiv},
       eprint = {0909.2115},
 primaryClass = {astro-ph.SR},
       adsurl = {https://ui.adsabs.harvard.edu/abs/2009ApJ...705..272R}
}

@ARTICLE{2012ApJ...752..108S,
       author = {{Sekse}, D.~H. and {Rouppe van der Voort}, L. and {De Pontieu}, B.},
        title = "{Statistical Properties of the Disk Counterparts of Type II Spicules from Simultaneous Observations of Rapid Blueshifted Excursions in Ca II 8542 and H{\ensuremath{\alpha}}}",
      journal = {\apj},
         year = 2012,
        month = jun,
       volume = {752},
       number = {2},
          eid = {108},
        pages = {108},
          doi = {10.1088/0004-637X/752/2/108},
archivePrefix = {arXiv},
       eprint = {1204.2943},
 primaryClass = {astro-ph.SR},
       adsurl = {https://ui.adsabs.harvard.edu/abs/2012ApJ...752..108S}
}

@ARTICLE{2013ApJ...769...44S,
       author = {{Sekse}, D.~H. and {Rouppe van der Voort}, L. and {De Pontieu}, B. and {Scullion}, E.},
        title = "{Interplay of Three Kinds of Motion in the Disk Counterpart of Type II Spicules: Upflow, Transversal, and Torsional Motions}",
      journal = {\apj},
         year = 2013,
        month = may,
       volume = {769},
       number = {1},
          eid = {44},
        pages = {44},
          doi = {10.1088/0004-637X/769/1/44},
archivePrefix = {arXiv},
       eprint = {1304.2304},
 primaryClass = {astro-ph.SR},
       adsurl = {https://ui.adsabs.harvard.edu/abs/2013ApJ...769...44S}
}

@ARTICLE{1943ApJ....98..116S,
       author = {{Swings}, P.},
        title = "{Edl{\'e}n's Identification of the Coronal Lines with Forbidden Lines of Fe X, XI, XIII, XIV, XV; Ni XII, XIII, XV, XVI; Ca XII, XIII, XV; a X, XIV}",
      journal = {\apj},
         year = 1943,
        month = jul,
       volume = {98},
        pages = {116-128},
          doi = {10.1086/144550},
       adsurl = {https://ui.adsabs.harvard.edu/abs/1943ApJ....98..116S}
}

@article{10.1093/mnras/107.2.211,
    author = {Alfvén, Hannes and Lindblad, B.},
    title = {Granulation, Magneto-Hydrodynamic Waves, and the Heating of the Solar Corona},
    journal = {Monthly Notices of the Royal Astronomical Society},
    volume = {107},
    number = {2},
    pages = {211-219},
    year = {1947},
    month = {06},
    issn = {0035-8711},
    doi = {10.1093/mnras/107.2.211},
    url = {https://doi.org/10.1093/mnras/107.2.211},
    eprint = {https://academic.oup.com/mnras/article-pdf/107/2/211/8076935/mnras107-0211.pdf},
}

@ARTICLE{1961ApJ...134..347O,
       author = {{Osterbrock}, Donald E.},
        title = "{The Heating of the Solar Chromosphere, Plages, and Corona by Magnetohydrodynamic Waves.}",
      journal = {\apj},
         year = 1961,
        month = sep,
       volume = {134},
        pages = {347},
          doi = {10.1086/147165},
       adsurl = {https://ui.adsabs.harvard.edu/abs/1961ApJ...134..347O}
}

@ARTICLE{1981JGR....8611463H,
       author = {{Hollweg}, J.~V. and {Isenberg}, P.~A.},
        title = "{On rotational forces in the solar wind}",
      journal = {\jgr},
         year = 1981,
        month = dec,
       volume = {86},
       number = {A13},
        pages = {11463-11463},
          doi = {10.1029/JA086iA13p11463},
       adsurl = {https://ui.adsabs.harvard.edu/abs/1981JGR....8611463H}
}

@ARTICLE{2009ApJ...701L...1D,
       author = {{De Pontieu}, Bart and {McIntosh}, Scott W. and {Hansteen}, Viggo H. and {Schrijver}, Carolus J.},
        title = "{Observing the Roots of Solar Coronal Heating{\textemdash}in the Chromosphere}",
      journal = {\apjl},
         year = 2009,
        month = aug,
       volume = {701},
       number = {1},
        pages = {L1-L6},
          doi = {10.1088/0004-637X/701/1/L1},
archivePrefix = {arXiv},
       eprint = {0906.5434},
 primaryClass = {astro-ph.SR},
       adsurl = {https://ui.adsabs.harvard.edu/abs/2009ApJ...701L...1D}
}

@ARTICLE{1982ApJ...255..743A,
       author = {{Athay}, R.~G. and {Holzer}, T.~E.},
        title = "{The role of spicules in heating the solar atmosphere}",
      journal = {\apj},
         year = 1982,
        month = apr,
       volume = {255},
        pages = {743-752},
          doi = {10.1086/159873},
       adsurl = {https://ui.adsabs.harvard.edu/abs/1982ApJ...255..743A}
}

@ARTICLE{2011ApJ...736L..24O,
       author = {{Okamoto}, Takenori J. and {De Pontieu}, Bart},
        title = "{Propagating Waves Along Spicules}",
      journal = {\apjl},
         year = 2011,
        month = aug,
       volume = {736},
       number = {2},
          eid = {L24},
        pages = {L24},
          doi = {10.1088/2041-8205/736/2/L24},
archivePrefix = {arXiv},
       eprint = {1106.4270},
 primaryClass = {astro-ph.SR},
       adsurl = {https://ui.adsabs.harvard.edu/abs/2011ApJ...736L..24O}
}

@ARTICLE{2011Natur.475..477M,
       author = {{McIntosh}, Scott W. and {de Pontieu}, Bart and {Carlsson}, Mats and {Hansteen}, Viggo and {Boerner}, Paul and {Goossens}, Marcel},
        title = "{Alfv{\'e}nic waves with sufficient energy to power the quiet solar corona and fast solar wind}",
      journal = {\nat},
         year = 2011,
        month = jul,
       volume = {475},
       number = {7357},
        pages = {477-480},
          doi = {10.1038/nature10235},
       adsurl = {https://ui.adsabs.harvard.edu/abs/2011Natur.475..477M}
}

@ARTICLE{2012ApJ...753..161M,
       author = {{Mart{\'\i}nez-Sykora}, Juan and {De Pontieu}, Bart and {Hansteen}, Viggo},
        title = "{Two-dimensional Radiative Magnetohydrodynamic Simulations of the Importance of Partial Ionization in the Chromosphere}",
      journal = {\apj},
         year = 2012,
        month = jul,
       volume = {753},
       number = {2},
          eid = {161},
        pages = {161},
          doi = {10.1088/0004-637X/753/2/161},
archivePrefix = {arXiv},
       eprint = {1204.5991},
 primaryClass = {astro-ph.SR},
       adsurl = {https://ui.adsabs.harvard.edu/abs/2012ApJ...753..161M}
}

@ARTICLE{2017Sci...356.1269M,
       author = {{Mart{\'\i}nez-Sykora}, J. and {De Pontieu}, B. and {Hansteen}, V.~H. and {Rouppe van der Voort}, L. and {Carlsson}, M. and {Pereira}, T.~M.~D.},
        title = "{On the generation of solar spicules and Alfv{\'e}nic waves}",
      journal = {Science},
         year = 2017,
        month = jun,
       volume = {356},
       number = {6344},
        pages = {1269-1272},
          doi = {10.1126/science.aah5412},
archivePrefix = {arXiv},
       eprint = {1710.07559},
 primaryClass = {astro-ph.SR},
       adsurl = {https://ui.adsabs.harvard.edu/abs/2017Sci...356.1269M}
}

@INPROCEEDINGS{2003SPIE.4853..341S,
       author = {{Scharmer}, Goran B. and {Bjelksjo}, Klas and {Korhonen}, Tapio K. and {Lindberg}, Bo and {Petterson}, Bertil},
        title = "{The 1-meter Swedish solar telescope}",
    booktitle = {Innovative Telescopes and Instrumentation for Solar Astrophysics},
         year = 2003,
       editor = {{Keil}, Stephen L. and {Avakyan}, Sergey V.},
       series = {Society of Photo-Optical Instrumentation Engineers (SPIE) Conference Series},
       volume = {4853},
        month = feb,
        pages = {341-350},
          doi = {10.1117/12.460377},
       adsurl = {https://ui.adsabs.harvard.edu/abs/2003SPIE.4853..341S}
}

@ARTICLE{2008ApJ...689L..69S,
       author = {{Scharmer}, G.~B. and {Narayan}, G. and {Hillberg}, T. and {de la Cruz Rodriguez}, J. and {L{\"o}fdahl}, M.~G. and {Kiselman}, D. and {S{\"u}tterlin}, P. and {van Noort}, M. and {Lagg}, A.},
        title = "{CRISP Spectropolarimetric Imaging of Penumbral Fine Structure}",
      journal = {\apjl},
         year = 2008,
        month = dec,
       volume = {689},
       number = {1},
        pages = {L69},
          doi = {10.1086/595744},
archivePrefix = {arXiv},
       eprint = {0806.1638},
 primaryClass = {astro-ph},
       adsurl = {https://ui.adsabs.harvard.edu/abs/2008ApJ...689L..69S}
}

@ARTICLE{2005SoPh..228..191V,
       author = {{Van Noort}, Michiel and {Rouppe Van Der Voort}, Luc and {L{\"o}fdahl}, Mats G.},
        title = "{Solar Image Restoration By Use Of Multi-frame Blind De-convolution With Multiple Objects And Phase Diversity}",
      journal = {\solphys},
         year = 2005,
        month = may,
       volume = {228},
       number = {1-2},
        pages = {191-215},
          doi = {10.1007/s11207-005-5782-z},
       adsurl = {https://ui.adsabs.harvard.edu/abs/2005SoPh..228..191V}
}

@ARTICLE{2012ApJ...750...22V,
       author = {{Vissers}, Gregal and {Rouppe van der Voort}, Luc},
        title = "{Flocculent Flows in the Chromospheric Canopy of a Sunspot}",
      journal = {\apj},
         year = 2012,
        month = may,
       volume = {750},
       number = {1},
          eid = {22},
        pages = {22},
          doi = {10.1088/0004-637X/750/1/22},
archivePrefix = {arXiv},
       eprint = {1202.5453},
 primaryClass = {astro-ph.SR},
       adsurl = {https://ui.adsabs.harvard.edu/abs/2012ApJ...750...22V}
}

@ARTICLE{2012SoPh..275...17L,
       author = {{Lemen}, James R. and {Title}, Alan M. and {Akin}, David J. and {Boerner}, Paul F. and {Chou}, Catherine and {Drake}, Jerry F. and {Duncan}, Dexter W. and {Edwards}, Christopher G. and {Friedlaender}, Frank M. and {Heyman}, Gary F. and {Hurlburt}, Neal E. and {Katz}, Noah L. and {Kushner}, Gary D. and {Levay}, Michael and {Lindgren}, Russell W. and {Mathur}, Dnyanesh P. and {McFeaters}, Edward L. and {Mitchell}, Sarah and {Rehse}, Roger A. and {Schrijver}, Carolus J. and {Springer}, Larry A. and {Stern}, Robert A. and {Tarbell}, Theodore D. and {Wuelser}, Jean-Pierre and {Wolfson}, C. Jacob and {Yanari}, Carl and {Bookbinder}, Jay A. and {Cheimets}, Peter N. and {Caldwell}, David and {Deluca}, Edward E. and {Gates}, Richard and {Golub}, Leon and {Park}, Sang and {Podgorski}, William A. and {Bush}, Rock I. and {Scherrer}, Philip H. and {Gummin}, Mark A. and {Smith}, Peter and {Auker}, Gary and {Jerram}, Paul and {Pool}, Peter and {Soufli}, Regina and {Windt}, David L. and {Beardsley}, Sarah and {Clapp}, Matthew and {Lang}, James and {Waltham}, Nicholas},
        title = "{The Atmospheric Imaging Assembly (AIA) on the Solar Dynamics Observatory (SDO)}",
      journal = {\solphys},
         year = 2012,
        month = jan,
       volume = {275},
       number = {1-2},
        pages = {17-40},
          doi = {10.1007/s11207-011-9776-8},
       adsurl = {https://ui.adsabs.harvard.edu/abs/2012SoPh..275...17L}
}

@ARTICLE{2012A&A...539A.146H,
       author = {{Hannah}, I.~G. and {Kontar}, E.~P.},
        title = "{Differential emission measures from the regularized inversion of Hinode and SDO data}",
      journal = {\aap},
         year = 2012,
        month = mar,
       volume = {539},
          eid = {A146},
        pages = {A146},
          doi = {10.1051/0004-6361/201117576},
archivePrefix = {arXiv},
       eprint = {1201.2642},
 primaryClass = {astro-ph.SR},
       adsurl = {https://ui.adsabs.harvard.edu/abs/2012A&A...539A.146H}
}

@ARTICLE{1997A&AS..125..149D,
       author = {{Dere}, K.~P. and {Landi}, E. and {Mason}, H.~E. and {Monsignori Fossi}, B.~C. and {Young}, P.~R.},
        title = "{CHIANTI - an atomic database for emission lines}",
      journal = {\aaps},
         year = 1997,
        month = oct,
       volume = {125},
        pages = {149-173},
          doi = {10.1051/aas:1997368},
       adsurl = {https://ui.adsabs.harvard.edu/abs/1997A&AS..125..149D}
}

@INPROCEEDINGS{1999ASSL..243...61L,
       author = {{Landi Degl'Innocenti}, E.},
        title = "{Evidence for ground-level atomic polarization in the solar atmosphere}",
    booktitle = {Polarization},
         year = 1999,
       editor = {{Nagendra}, K.~N. and {Stenflo}, J.~O.},
       series = {Astrophysics and Space Science Library},
       volume = {243},
        month = jan,
        pages = {61-71},
          doi = {10.1007/978-94-015-9329-8_5},
       adsurl = {https://ui.adsabs.harvard.edu/abs/1999ASSL..243...61L}
}

@ARTICLE{2004SoPh..225..293K,
       author = {{Kontar}, Eduard P. and {Piana}, Michele and {Massone}, Anna Maria and {Emslie}, A. Gordon and {Brown}, John C.},
        title = "{Generalized Regularization Techniques with Constraints for the Analysis of Solar Bremsstrahlung X-ray Spectra}",
      journal = {\solphys},
         year = 2004,
        month = dec,
       volume = {225},
       number = {2},
        pages = {293-309},
          doi = {10.1007/s11207-004-4140-x},
archivePrefix = {arXiv},
       eprint = {astro-ph/0409688},
 primaryClass = {astro-ph},
       adsurl = {https://ui.adsabs.harvard.edu/abs/2004SoPh..225..293K}
}

@ARTICLE{2013ApJ...764..164S,
       author = {{Sekse}, D.~H. and {Rouppe van der Voort}, L. and {De Pontieu}, B.},
        title = "{On the Temporal Evolution of the Disk Counterpart of Type II Spicules in the Quiet Sun}",
      journal = {\apj},
         year = 2013,
        month = feb,
       volume = {764},
       number = {2},
          eid = {164},
        pages = {164},
          doi = {10.1088/0004-637X/764/2/164},
archivePrefix = {arXiv},
       eprint = {1212.4988},
 primaryClass = {astro-ph.SR},
       adsurl = {https://ui.adsabs.harvard.edu/abs/2013ApJ...764..164S}
}

@ARTICLE{2014ApJ...792L..15P,
       author = {{Pereira}, T.~M.~D. and {De Pontieu}, B. and {Carlsson}, M. and {Hansteen}, V. and {Tarbell}, T.~D. and {Lemen}, J. and {Title}, A. and {Boerner}, P. and {Hurlburt}, N. and {W{\"u}lser}, J.~P. and {Mart{\'\i}nez-Sykora}, J. and {Kleint}, L. and {Golub}, L. and {McKillop}, S. and {Reeves}, K.~K. and {Saar}, S. and {Testa}, P. and {Tian}, H. and {Jaeggli}, S. and {Kankelborg}, C.},
        title = "{An Interface Region Imaging Spectrograph First View on Solar Spicules}",
      journal = {\apjl},
         year = 2014,
        month = sep,
       volume = {792},
       number = {1},
          eid = {L15},
        pages = {L15},
          doi = {10.1088/2041-8205/792/1/L15},
archivePrefix = {arXiv},
       eprint = {1407.6360},
 primaryClass = {astro-ph.SR},
       adsurl = {https://ui.adsabs.harvard.edu/abs/2014ApJ...792L..15P}
}

@ARTICLE{2015ApJ...806..170S,
       author = {{Skogsrud}, H. and {Rouppe van der Voort}, L. and {De Pontieu}, B. and {Pereira}, T.~M.~D.},
        title = "{On the Temporal Evolution of Spicules Observed with IRIS, SDO, and Hinode}",
      journal = {\apj},
         year = 2015,
        month = jun,
       volume = {806},
       number = {2},
          eid = {170},
        pages = {170},
          doi = {10.1088/0004-637X/806/2/170},
archivePrefix = {arXiv},
       eprint = {1505.02525},
 primaryClass = {astro-ph.SR},
       adsurl = {https://ui.adsabs.harvard.edu/abs/2015ApJ...806..170S}
}

@ARTICLE{2022MNRAS.509.5523N,
       author = {{Nived}, V.~N. and {Scullion}, E. and {Doyle}, J.~G. and {Susino}, R. and {Antolin}, P. and {Spadaro}, D. and {Sasso}, C. and {Sahin}, S. and {Mathioudakis}, M.},
        title = "{Implications of spicule activity on coronal loop heating and catastrophic cooling}",
      journal = {\mnras},
         year = 2022,
        month = feb,
       volume = {509},
       number = {4},
        pages = {5523-5537},
          doi = {10.1093/mnras/stab3277},
archivePrefix = {arXiv},
       eprint = {2111.07967},
 primaryClass = {astro-ph.SR},
       adsurl = {https://ui.adsabs.harvard.edu/abs/2022MNRAS.509.5523N}
}

% Alternatively you could enter them by hand, like this:
% This method is tedious and prone to error if you have lots of references
%\begin{thebibliography}{99}
%\bibitem[\protect\citeauthoryear{Author}{2012}]{Author2012}
%Author A.~N., 2013, Journal of Improbable Astronomy, 1, 1
%\bibitem[\protect\citeauthoryear{Others}{2013}]{Others2013}
%Others S., 2012, Journal of Interesting Stuff, 17, 198
%\end{thebibliography}

%%%%%%%%%%%%%%%%%%%%%%%%%%%%%%%%%%%%%%%%%%%%%%%%%%

%%%%%%%%%%%%%%%%% APPENDICES %%%%%%%%%%%%%%%%%%%%%

%%%%%%%%%%%%%%%%%%%%%%%%%%%%%%%%%%%%%%%%%%%%%%%%%%

% Don't change these lines
\bsp	% typesetting comment
\label{lastpage}
\end{document}